\documentclass{article}
\usepackage[OT1]{fontenc} 
\usepackage{spconf,amsmath,amssymb,amsfonts,graphicx}
\usepackage{cite}
\usepackage{mathrsfs}
\usepackage{algorithm}
\usepackage{algorithmic}
\usepackage{enumerate}
\usepackage{url}
\usepackage{multirow}
\usepackage{fancyhdr}
\usepackage{color}
\usepackage{etoolbox}
\patchcmd\thebibliography
 {\labelsep}
 {\labelsep\itemsep=-1pt\relax}
 {}
 {\typeout{Couldn't patch the command}}
\title{Low-rankness of Complex-valued Spectrogram \\ and Its Application to Phase-aware Audio Processing}
\name{Yoshiki Masuyama, Kohei Yatabe and Yasuhiro Oikawa \vspace{-4pt}}
\address{\fontsize{11pt}{0pt}\selectfont Department of Intermedia Art and Science, Waseda University, Tokyo, Japan}
\begin{document}
\ninept
\maketitle
\begin{abstract}
Low-rankness of amplitude spectrograms has been effectively utilized in audio signal processing methods including non-negative matrix factorization.
However, such methods have a fundamental limitation owing to their amplitude-only treatment where the phase of the observed signal is utilized for resynthesizing the estimated signal.
In order to address this limitation, we directly treat a complex-valued spectrogram and show a complex-valued spectrogram of a sum of sinusoids can be approximately low-rank by modifying its phase.
For evaluating the applicability of the proposed low-rank representation, we further propose a convex prior emphasizing harmonic signals, and it is applied to audio denoising.
\end{abstract}

%
\vspace{-2pt}
\section{Introduction}
\vspace{-2pt}

In audio signal processing, low-rank representation of amplitude spectrograms has been utilized extensively \cite{ISNMF, MNMF, NMFspeech1, RPCA, NMFED, ILRMA, Yatabe}.
For instance, the non-negative matrix factorization (NMF) has been developed in many applications including source separation \cite{NMFsep}, audio inpainting \cite{NMFinp}, and music transcription \cite{NMFtrans}.
While those methods have successfully applied to various problems, they have a limitation owing to their amplitude-only treatment.

Recent studies have shown the importance of phase \cite{Phase0, Phase1, Phase2}, and some phase-aware extensions of NMF have been proposed \cite{CNMF, TSF}.
In addition to amplitude spectrograms, the complex NMF (CNMF) treats phase at each time-frequency bin as the independent variables to be optimized \cite{CNMF}.
Meanwhile, the time-domain spectrogram factorization (TSF) implements NMF-like signal decomposition in the time domain for implicitly considering phase based on the consistency of a spectrogram \cite{TSF}.
Although these methods include phase in their models, low-rankness is only imposed on amplitude spectrograms, and the explicit structure of the phase was not considered.

A very recent study revealed the relation between the rank and phase of a complex-valued spectrogram \cite{BeingLR}.
The phase of a sum of sinusoids has a distinctive structure which has been widely utilized in audio signal processing \cite{PCTV, PU1, iPCTV, PU2, MasIWAENC, HPSS}, and the rank of its complex-valued spectrogram depends on the number of the sinusoids \cite{BeingLR}.
This theoretical result indicates that the rank of the complex-valued spectrogram increases as the number of sinusoids increases, while its amplitude can stay low-rank.
Therefore, imposing low-rankness only on amplitude spectrograms as in CNMF and TSF seems to be justified, and low-rank treatment of a complex-valued spectrogram sounds inappropriate.
However, we found the complex-valued spectrogram can be well approximated by a low-rank matrix applying a specific modification of phase.

In this paper, we propose a low-rank representation of a complex-valued spectrogram by applying the instantaneous phase correction introduced in \cite{iPCTV}.
This phase modification is based on the phase model of harmonic signals, and the phase corrected complex-valued spectrogram of harmonic signals becomes as low-rank as its amplitude spectrogram with some assumptions.
As an example of the applications of the proposed low-rank representation, we further propose a convex prior which emphasizes sinusoidal components through the low-rankness, and its effectiveness is demonstrated by an audio denoising experiment.

\vspace{-2pt}
\section{Previous work}
\label{sec: 2}
\vspace{-2pt}

Let the short-time Fourier transform (STFT) of a signal $\mathbf{x} = [{x}[0], \ldots ,{x}[L-1]]^\top \in \mathbb{R}^{L}$ with a window $\mathbf{w} \in \mathbb{R}^{L}$ be
\begin{align}
\mathcal{G}^\mathbf{w} (\mathbf{x})[\xi, \tau] = \sum_{l = 0}^{L-1} {x}[l+a\tau] {w}[l] e^{-2\pi j \xi b l / L},
\end{align}
where $j = \sqrt{-1}$, $a$ and  $b$ are the time and frequency shifting steps, $\tau = 0, 1, \ldots, T-1$ and $\xi = 0, 1, \ldots, K-1$ denote the time and frequency indices, and index overflow is treated by zero-padding.
This STFT can be represented by a matrix form:
\begin{align}
\mathcal{G}^\mathbf{w} (\mathbf{x}) &= \mathbf{F} \mathrm{diag}(\mathbf{w}) \mathbf{X}, \\
\mathbf{X} &= [ {\mathbf{x}}_0, \mathbf{x}_1, \ldots, \mathbf{x}_{T-1} ], \label{eq:hankel}
\end{align}
where $\mathbf{X}$ is the horizontal concatenation of $T$ patches of the signal $\mathbf{x}_{\tau} = [ x[a\tau], x[a\tau+1], \ldots, x[a\tau+L-1] ]^\top$, $\mathbf{z}^\top$ is the transpose of $\mathbf{z}$, $\mathbf{F}$ is the discrete Fourier transform matrix, and $\mathrm{diag}(\mathbf{w})$ denotes the diagonal matrix whose diagonal elements are given by $\mathbf{w}$.

Let a sinusoid be written as
\begin{equation}
s_0[l] = A_0 e^{2 \pi j f_0 l /L + \phi_0},
\end{equation}
where $A_0 \in \mathbb{R}_+ $, $f_0 \in [0, L/2)$ and $\phi_0 \in [0, 2\pi)$ are the amplitude, frequency, and initial phase of the sinusoid, respectively.
Considering patches of this sinusoid $\mathbf{s}_{0}^0,\mathbf{s}_{1}^0, \ldots, \mathbf{s}_{T-1}^0$ given by
\begin{equation}
\mathbf{s}_{\tau}^0 = [ s_0[a\tau], s_0[a\tau+1], \ldots, s_0[a\tau+L-1] ]^\top,
\end{equation}
every patch has the following relation:
\begin{equation}
{\mathbf{s}}_{\tau}^0 =   e^{2 \pi j f a / L} {\mathbf{s}}_{\tau-1}^0 = e^{2 \pi j f a \tau / L}{\mathbf{s}}_{0}^0.
\label{relation1}
\end{equation}
Considering the matrix $\mathbf{S}_0$ whose columns are given by ${\mathbf{s}}_{\tau}^0$ as Eq.~\eqref{eq:hankel}, $\mathrm{rank}(\mathbf{S}_0) = 1$, where $\mathrm{rank}(\mathbf{S}_0)$ is the rank of the matrix $\mathbf{S}_0$.
This is because all columns of $\mathbf{S}_0$ are given by a complex-valued scalar multiplication of $\mathbf{s}_0$.
Since $\mathrm{rank}( \mathcal{G}^{\mathbf{w}}(\mathbf{x}) )$ coincides with $\mathrm{rank}( \mathbf{X} )$ when $\mathrm{diag} (\mathbf{w})$ is full rank (i.e., $w[l] \neq 0$\, $\forall l$) by the unitarity of $\mathbf{F}$, the rank of the complex-valued spectrogram of a sinusoid is $1$ as described in \cite{BeingLR}.

The previous work also showed that the rank of a complex-valued spectrogram of a sum of sinusoids depends on the number of sinusoids \cite{BeingLR}, while its amplitude can be well approximated by a rank-$1$ matrix regardless of the number of sinusoids.
Consider a sum of $H$ sinusoids:
\begin{equation}
s[l] = \sum_{h=0}^{H-1} s_h[l] = \sum_{h=0}^{H-1} {A}_h e^{2 \pi j f_h l/L + \phi_h},
\label{sumsin}
\end{equation}
where ${A}_h\in \mathbb{R}_+$, $f_h \in [0, L/2)$ $(f_p \neq f_q$ when $p \neq q$), and $\phi_h \in [0, 2\pi)$ are the amplitude, frequency, and initial phase of $h$th sinusoid, respectively.
The matrix which contains its patches is defined as $\mathbf{S} = \sum_{h=0}^{H-1} \mathbf{S}_h$, and $\mathrm{rank}(\mathcal{G}^\mathbf{w} (\mathbf{s}))$ increases as the number of sinusoids $H$ increases%
\footnote{
As described in \cite{BeingLR}, the complex-valued spectrogram of a sum of $H$ sinusoids becomes a rank-$H$ matrix when the frequencies of sinusoids are on the discrete Fourier grid.
Note that \cite{BeingLR} considers STFT with maximal redundancy and periodic extension of the signal and window.
An exact characterization for more general cases was not presented.
}.

Low-rank representation of an amplitude spectrogram has been well-accepted in audio signal processing since an amplitude spectrogram of a sum of sinusoids is low-rank regardless of the number of sinusoids.
In contrast, the low-rankness of a complex-valued spectrogram has not been considered owing to the above nature.

\vspace{-2pt}
\section{Proposed low-rank representation of complex-valued spectrograms}
\label{sec: 3}
\vspace{-2pt}

In this section, we show that a complex-valued spectrogram of a sum of sinusoids becomes low-rank when the instantaneous phase correction \cite{iPCTV} is applied.
The characteristic of the complex-valued spectrogram with this phase correction is reviewed first, and then its low-rankness is validated with some numerical examples.

\vspace{-2pt}
\subsection{Instantaneous phase corrected STFT (iPC-STFT) \cite{iPCTV}}
\label{sec:iPCSTFT}
\vspace{-2pt}

\begin{figure}[t]
	\centering
	\includegraphics[width=0.99\columnwidth]{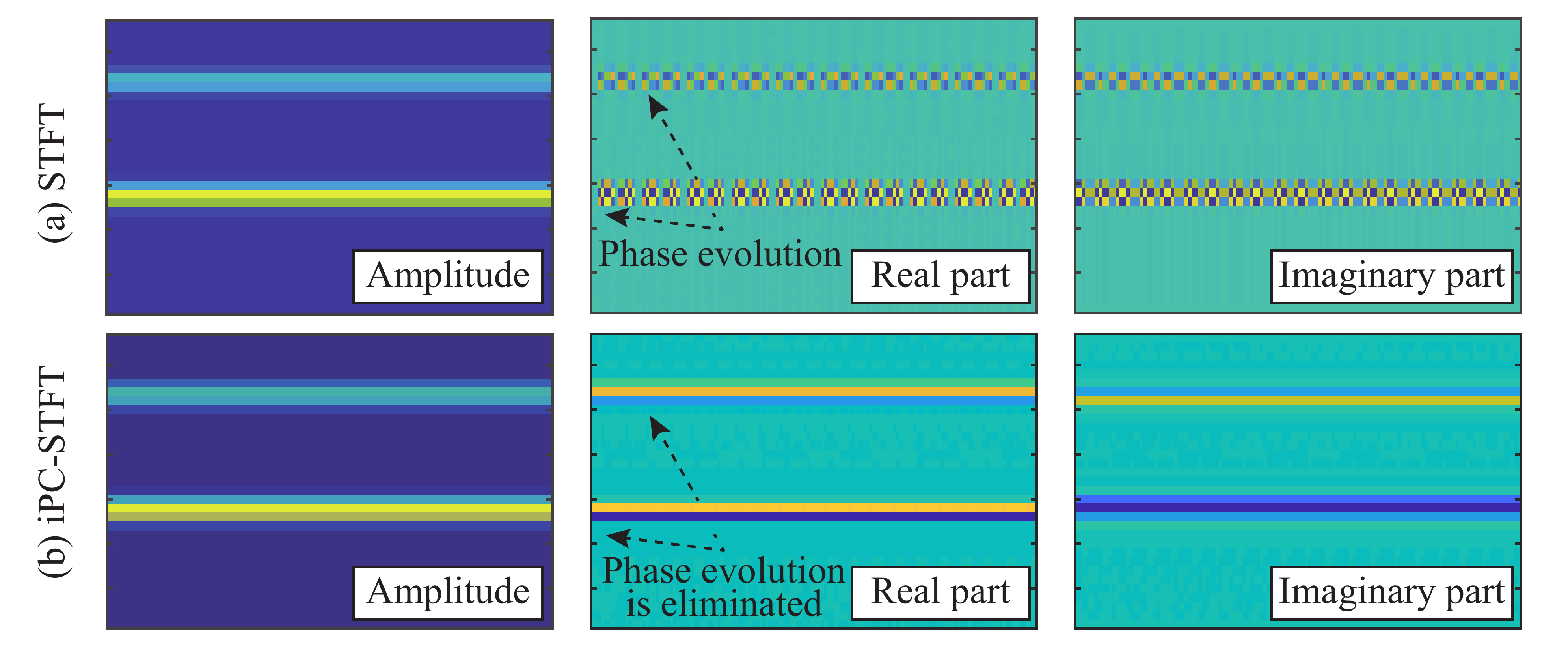}
	\vspace{-10pt}
	\caption{Comparison of spectrograms of two sinusoids calculated by (a) the usual STFT and (b) iPC-STFT.}
	\label{fig:illust}
	\vspace{-8pt}
\end{figure}

As in the previous section, the rank of a complex-valued spectrogram of a sum of sinusoids depends on the number of sinusoids, while its amplitude is well represented by a rank-$1$ matrix.
This is because the evolution of the phase is different for each sinusoid as illustrated in Fig.~\ref{fig:illust}(a), where a sum of two sinusoids is shown in the time-frequency domain.
Since the phase evolves along time, the real and imaginary parts of the complex-valued spectrogram periodically fluctuate.
Their periods are different in each sinusoid, which results in the increase of the rank.
If this phase evolution is eliminated, the rank of the complex-valued spectrogram can be reduced.

The phase evolution is closely related to the instantaneous frequency at each time-frequency bin.
When we assume each sinusoid in Eq.~\eqref{sumsin} is sufficiently separated (i.e., the interference from other sinusoids can be ignored in the region dominated by a sinusoid), the complex-valued spectrogram of the sum of sinusoids has the following relation by utilizing the instantaneous frequency \cite{MasIWAENC}:
\begin{equation}
\mathcal{G}^\mathbf{w} (\mathbf{s}) [\xi, \tau+1] = e^{2 \pi j v[\xi, \tau] a/L} \mathcal{G}^\mathbf{w} (\mathbf{s})[\xi, \tau],
\label{eq:relationSTFT}
\end{equation}
where $v[\xi, \tau]$ is different for each sub-band but same in all time-frames.
That is, the phase of the complex-valued spectrogram evolves as a constant multiple of the instantaneous frequency $v[\xi, \tau]$.

To cancel this phase evolution, the instantaneous phase corrected STFT (iPC-STFT) was proposed as follows: \cite{iPCTV}
\begin{equation}
\mathcal{G}^\mathbf{w}_\mathrm{iPC}(\mathbf{x}) = \mathbf{E} \odot \mathcal{G}^\mathbf{w}(\mathbf{x}),
\label{eq: defiPCSTFT}
\end{equation}
where $\odot$ is the Hadamard product, $\mathbf{E}$ is the instantaneous phase correction matrix whose element is defined by
\begin{equation}
E[\xi, \tau] = \prod_{\eta=0}^{\tau-1} e^{-2 \pi j v[\xi, \eta] a/L},
\label{eq: ipc}
\end{equation}
and $E[\xi, 0] = 1$ for all $\xi$.
This matrix cancels the phase evolution in Eq.~\eqref{eq:relationSTFT}, which results in the low-rank complex-valued spectrogram as shown in Fig.~\ref{fig:illust}(b).
Both real and imaginary parts of the complex-valued spectrogram are constant at each sub-band, and therefore its rank is reduced.
Note that this correction of phase can be easily inverted by multiplying the complex conjugate of $\mathbf{E}$.

In reality, the instantaneous frequency $v[\xi,\tau]$ is not known and must be estimated from the observed signal.
One simple method is to directly calculate the time-differential of phase:
\begin{equation}
\vspace{-2pt}
v[\xi,\tau] = b\xi - \mathrm{Im} \biggl[ \frac{\mathcal{G}^\mathbf{w'}(\mathbf{x}) [\xi, \tau]}{\mathcal{G}^\mathbf{w} (\mathbf{x}) [\xi,\tau]} \biggr],
\label{eq:reasign}
\vspace{-2pt}
\end{equation}
where $\mathbf{w}'$ is time-derivative of the window $\mathbf{w}$ \cite{Reasign1,Reasign3,Reasign4}, and $\mathrm{Im}[z]$ is the imaginary part of $z$.
Once the instantaneous frequency is estimated, the instantaneous phase correction is uniquely defined as an invertible linear transform in the proposed denoising scheme introduced in Section~\ref{sec: ipclr}.
Note that iPC-STFT can be applied to arbitrary signal which may not consist of pure sinusoids.

\vspace{-2pt}
\subsection{Low-rankness of iPC-STFT spectrogram}
\vspace{-2pt}

The previous work \cite{iPCTV} only considered time-directional smoothness of iPC-STFT, and no further characteristics have been shown.
In this paper, we show the low-rankness of the complex-valued spectrogram calculated by iPC-STFT.
Such low-rankness in the complex domain should be important for extending ordinary studies of low-rank audio modeling.

From Eqs.~\eqref{eq:relationSTFT}--\eqref{eq: ipc}, with some conditions mentioned in the previous subsection, the following neighborhood relation can be obtained for iPC-STFT of a sum of sinusoids:
\begin{align}
\mathcal{G}^\mathbf{w}_{\mathrm{iPC}} (\mathbf{s})[\xi, \tau+1] &= \prod_{\eta=0}^{\tau} e^{-2 \pi j v[\xi, \eta] a/L} \mathcal{G}^\mathbf{w} (\mathbf{s})[\xi, \tau+1], \nonumber \\
&= \prod_{\eta=0}^{\tau-1} e^{-2 \pi j v[\xi, \eta] a/L} \mathcal{G}^\mathbf{w} (\mathbf{s})[\xi, \tau], \nonumber \\
&= \mathcal{G}^\mathbf{w} (\mathbf{s})[\xi, 0] \quad\left( = \mathcal{G}^\mathbf{w}_{\mathrm{iPC}} (\mathbf{s})[\xi, 0] \right).
\label{eq: iPCtrans}
\end{align}
That is, all columns of $\mathcal{G}^\mathbf{w}_{\mathrm{iPC}}(\mathbf{s})$ is equivalent to the first column of $\mathcal{G}^\mathbf{w}_{\mathrm{iPC}}(\mathbf{s})$ when the instantaneous phase correction eliminates the phase evolution completely (i.e., when the instantaneous frequency is accurately estimated).
Hence, $\mathrm{rank}( \mathcal{G}^\mathbf{w}_{\mathrm{iPC}}(\mathbf{s}) ) = 1$ regardless of the number of sinusoids.

The above relation of iPC-STFT can be generalized to a signal beyond a sum of sinusoids.
The relation in Eq.~\eqref{eq:relationSTFT} can be understood as the well-accepted sinusoidal model \cite{PU1}:
\begin{equation}
\phi[\xi, \tau+1] = \phi[\xi, \tau] + 2 \pi v[\xi, \tau] a/L,
\label{eq:sinusoidalmodel}
\end{equation}
where $\phi$ is a phase spectrogram.
This equation indicates that $\phi[\xi, \tau] = \phi[\xi, 0] + \sum_{\eta=0}^{\tau-1} 2 \pi v[\xi, \eta] a/L$, and therefore
\begin{align}
\mathcal{G}^\mathbf{w}_{\mathrm{iPC}} (\mathbf{x})[\xi, \tau] &= \prod_{\eta=0}^{\tau-1} e^{-2 \pi j v[\xi, \eta] a/L} \mathcal{G}^\mathbf{w} (\mathbf{x})[\xi, \tau], \nonumber \\
&= \prod_{\eta=0}^{\tau-1} e^{-2 \pi j v[\xi, \eta] a/L}  |\mathcal{G}^\mathbf{w} (\mathbf{x})[\xi, \tau]| e^{2 \pi j \phi[\xi, \tau] a/L}, \nonumber \\
&= |\mathcal{G}^\mathbf{w} (\mathbf{x})[\xi, \tau]|   e^{2 \pi j \phi[\xi, 0] a/L},
\label{eq: iPCtrans2}
\end{align}
i.e., the instantaneous phase correction converts a complex-valued spectrogram into a constant multiple of its amplitude.
Hence, $\mathrm{rank}(\mathcal{G}^\mathbf{w}_{\mathrm{iPC}} (\mathbf{x})) = \mathrm{rank}(|\mathcal{G}^\mathbf{w} (\mathbf{x})|)$ whenever the instantaneous frequency at each time-frequency bin $v[\xi, \tau]$ is estimated exactly.
When $v[\xi, \tau]$ can be estimated only approximately, the relation also becomes approximation: $\mathrm{rank}(\mathcal{G}^\mathbf{w}_{\mathrm{iPC}} (\mathbf{x})) \approx \mathrm{rank}(|\mathcal{G}^\mathbf{w} (\mathbf{x})|)$.

We stress that the low-rankness of a complex-valued spectrogram is completely different from that of the amplitude.
For instance, amplitude spectrogram of the Gaussian noise realized in the time domain can be well approximated by rank-$1$ because its energy is almost the same for all time-frequency bins.
In contrast, its complex-valued spectrogram is not low-rank because its phase dose not obey Eq.~\eqref{eq:sinusoidalmodel}.
Such difference between the low-rankness of amplitude and complex-valued spectrograms is experimentally illustrated in the next subsection.

\begin{figure}[t]
	\centering
	\includegraphics[width=0.99\columnwidth]{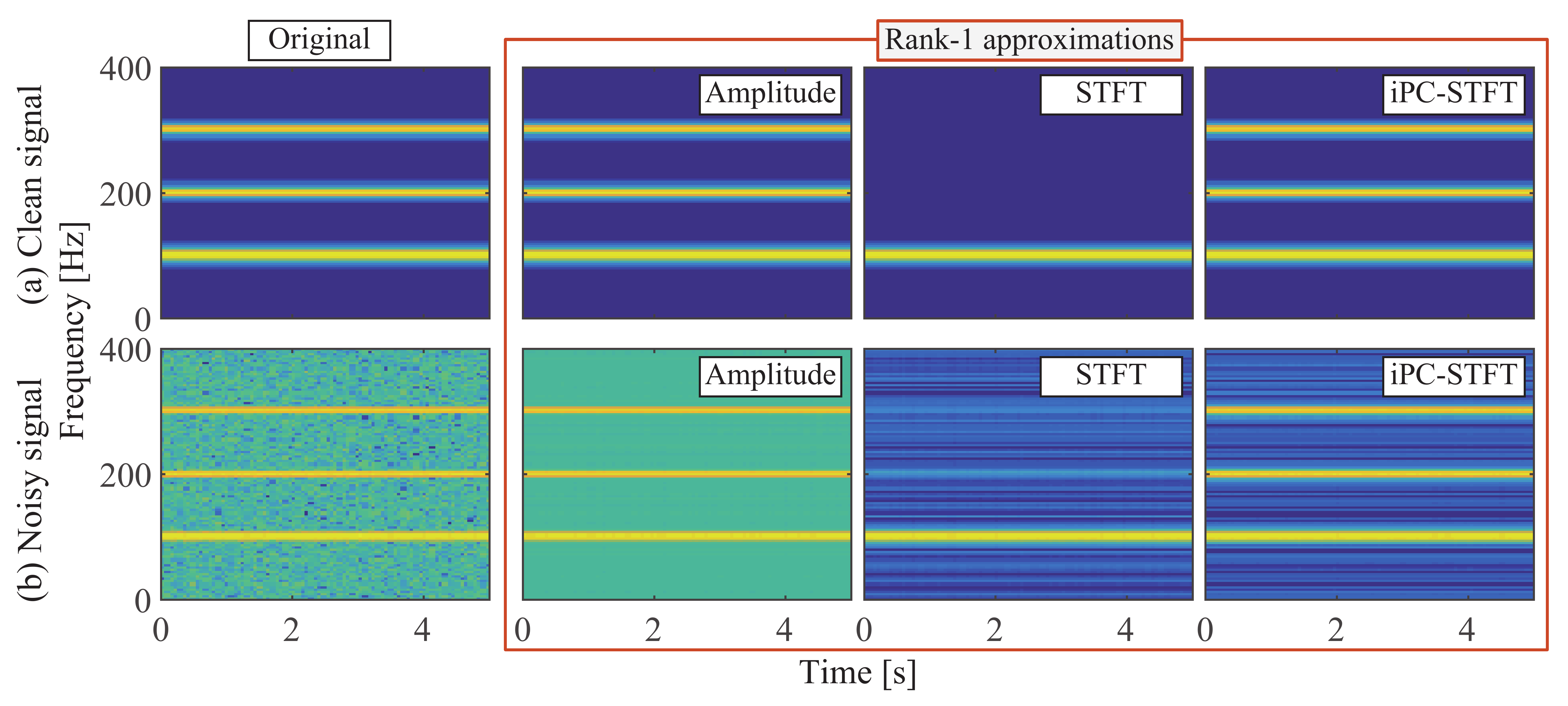}
	\vspace{-10pt}
	\caption{Amplitude of the rank-$1$ approximated spectrograms of a clean sum of three sinusoids and degraded one.}
	\label{fig:exam}
	\vspace{-8pt}
\end{figure}

\begin{table}[t]
\centering
\caption{SNR of rank-$1$ approximation of spectrograms in Fig.~\ref{fig:exam}.}
\vspace{2pt}
\label{tab:snr}
\scalebox{1.0}{
	\begin{tabular}{c|c|c|c|c|c}
		\hline
		 \multicolumn{2}{c|}{ } & \multicolumn{3}{c}{Input SNR [dB]} & \multicolumn{1}{|c}{ } \\
		 \hline
		 \multicolumn{1}{c|}{ } &\multicolumn{1}{c|}{shift size} & \multicolumn{1}{c|}{0} & \multicolumn{1}{c|}{10} & \multicolumn{1}{c|}{20} & \multicolumn{1}{|c}{Clean} \\ 
		 \hline
		 \multirow{3}{*}{Amplitude} & {1/2} & {1.3} & {11.3} & {21.4} &{64.4} \\ & {1/4} & {1.3} & {11.4} & {21.4} &{64.3} \\ & {1/8} & {1.3} & {11.4} & {21.4} & {62.9} \\
		 \hline
		 \multirow{3}{*}{STFT} & {1/2} & {2.2} & {2.3} & {2.3} & {2.3}\\ & {1/4} & {2.2} & {2.3} & {2.3}  & {2.3} \\ & {1/8} & {2.3} & {2.3} & {2.3} & {2.3} \\
		 \hline
		 \multirow{3}{*}{iPC-STFT} & {1/2} & {18.8} & {28.9} & {38.7} & {52.3} \\ & {1/4} & {21.8} & {31.6} & {41.5} & {55.4} \\ & {1/8} & {24.5} & {34.3} & {44.2} & {55.3} \\
 		 \hline
	\end{tabular}
	}
\vspace{-8pt}
\end{table}

\begin{figure}[t]
	\centering
	\includegraphics[width=0.99\columnwidth]{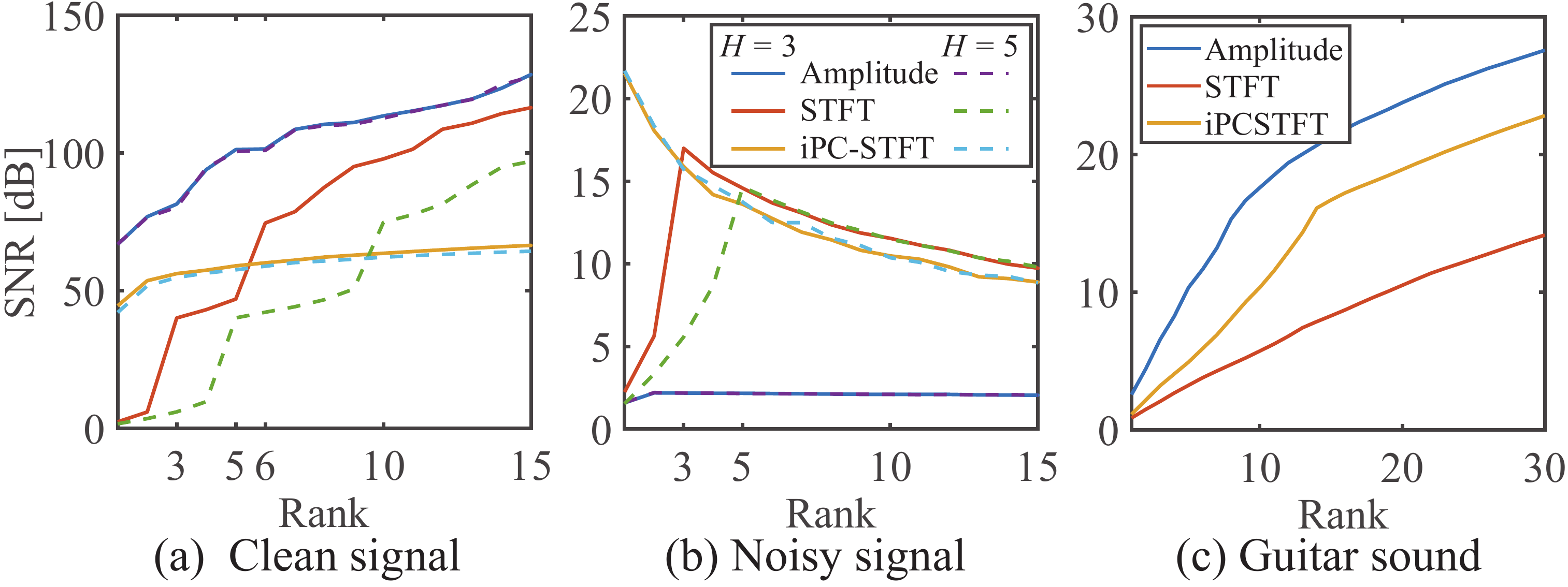}
	\vspace{-10pt}
	\caption{SNR of rank-$k$ approximation of spectrograms of (a) a sum of pure sinusoids, (b) (a) with the additive Gaussian noise, and (c) the guitar sound without noise.}
	\label{fig:SVD}
	\vspace{-8pt}
\end{figure}

\vspace{-2pt}
\subsection{Numerical examples}
\vspace{-2pt}

For illustrating the property of iPC-STFT and its low-rank representation, some simple examples are shown here.
Firstly, a sum of three sinusoids, $s[l] = \sum_{h=0}^{H-1} A_h \sin({2 \pi f_h l/L})$, is considered, where $H = 3$, $A_h = 10 - h$, $f_h = (h+1) f_0$, $f_0 = 100$ Hz, and the sampling frequency was $16000$ Hz.
STFT was calculated with the Hann window ($4096$ samples), and the instantaneous frequency was estimated by Eq.~\eqref{eq:reasign}.
Here the rank-$k$ approximations were calculated by the truncated singular value decomposition. 

Rank-1 approximations of the three sinusoids are illustrated in Fig.~\ref{fig:exam}.
The leftmost column is the original signals, and second to fourth columns represent the rank-1 approximations of the amplitude, complex (STFT), and complex (iPC-STFT) spectrograms, respectively.
From Fig.~\ref{fig:exam}(a), the usual complex-valued (STFT) spectrogram can only represent one sinusoid by the rank-$1$ approximation as described in Section~\ref{sec: 2}.
In contrast, both amplitude and iPC-STFT spectrograms can simultaneously represent all three sinusoids by rank-1 approximation.
This result confirms the relation $\mathrm{rank}(|\mathcal{G}^\mathbf{w} (\mathbf{x})|) \! \approx \! \mathrm{rank}(\mathcal{G}^\mathbf{w}_{\mathrm{iPC}} (\mathbf{x}))$ described in the previous subsection.
Fig.~\ref{fig:exam}(b) represents a noisy version of Fig.~\ref{fig:exam}(a), where the complex Gaussian noise was added in the time-frequency domain.
In this noisy case, the rank-$1$ approximation of the amplitude spectrogram stayed noisy because the amplitude spectrogram of stationary noise can be well approximated by a low-rank matrix.
On the other hand, rank-$1$ approximation of complex-valued (STFT and iPC-STFT) spectrograms can remove the Gaussian noise to some extent.
This is because iPC-STFT takes the phase structure of sinusoidal components into account, while the amplitude specrogram ignores the phase information.
That is, the proposed low-rank representation of complex-valued spectrograms can distinguish sinusoidal components from noise.
Some quantitative data of those rank-1 approximations are shown in Table~\ref{tab:snr}, where the signal-to-noise ratio (SNR) of the original signal and the shift size of window are varied.
It confirms that only iPC-STFT can improve SNR of Fig.~\ref{fig:exam}(b) by the rank-1 approximation%
\footnote{
For calculating SNR of the rank-$1$ approximation of amplitude spectrograms in the complex domain, the observed noisy phase was utilized.
This approach was often utilized in NMF for resynthesizing the estimated signal.
}.

Next, SNR of rank-$k$ approximations of sums of three or five sinusoids are shown in Fig.~\ref{fig:SVD}(a) and \ref{fig:SVD}(b).
The shift size of the window was $1/4$, and the input SNR was $10$ dB for Fig.~\ref{fig:SVD}(b).
The rank-$k$ approximation of the amplitude spectrogram resulted in lowest SNR because it does not distinguish sinusoidal components from the stationary noise. 
Since the rank of the usual complex-valued (STFT) spectrogram is decided by the number of sinusoidal components \cite{BeingLR}, SNR of STFT was low for the rank-1 approximation $(k=1)$ but significantly improved when the rank $k$ coincides with the number of sinusoids $H$.
For the noisy situation in Fig.~\ref{fig:SVD}(b), the highest SNR of STFT was achieved when at $k\!=\!H$.
In contrast, iPC-STFT obtained a low-rank complex-valued spectrograms which can be well approximated by a rank-1 matrix.
As a result, the highest SNR in Fig.~\ref{fig:SVD}(b) was achieved by the rank-1 approximation of the complex-valued iPC-STFT spectrogram regardless of the number of sinusoids $H$.
Finally, rank-$k$ approximation of a real guitar sound is illustrated in Fig.~\ref{fig:SVD}(c), where the guitar sample was obtained from IDMT-SMT-GUITAR database%
\footnote{\url{http://www.idmt.fraunhofer.de/en/business_units/smt/guitar.html}} \cite{Guitar}, and the sampling frequency was $44100$ Hz.
The complex-valued spectrogram calculated by iPC-STFT can represent the real guitar sound better than that of the usual STFT.
One reason of lower SNR of iPC-STFT comparing to the amplitude-only approximation should be the error of the instantaneous frequency estimation which can be improved by considering an estimation method more sophisticated than Eq.~\eqref{eq:reasign}.

\vspace{-2pt}
\section{Application}
\vspace{-2pt}

\vspace{-2pt}
\subsection{Proposed low-rank representation as a signal prior}
\label{sec: ipclr}
\vspace{-2pt}

For applying the proposed low-rank representation to audio signal processing,
we propose a novel prior, named instantaneous phase corrected low-rankness (iPCLR), defined by
\begin{equation}
\mathcal{P}_\mathrm{iPCLR}(\mathbf{x}) = \left\| \mathcal{G}^\mathbf{w}_{\mathrm{iPC}}(\mathbf{x}) \right\|_\ast,
\end{equation}
where $\left\|\;\cdot\;\right\|_\ast$ denotes the nuclear norm which is the convex envelope of the rank function $\mathrm{rank}(\cdot)$.
Thanks to the property of iPC-STFT illustrated in the previous section, this prior emphasizes sinusoidal components and suppresses non-sinusoidal components such as random noise.
Note that the instantaneous frequency is pre-computed and fixed to make iPC-STFT into a linear operator (independent of $\mathbf{x}$), and thus this prior is convex and can be easily incorporated with other priors.

As an application of iPCLR, the following convex optimization problem is considered for audio denoising,
\begin{equation}
\mathbf{x}^{\star} = \mathrm{arg}\min_{\mathbf{x}} \frac{1}{2} \left\| \mathbf{x} - \mathbf{d}\right\|_2^2 + \lambda \,\mathcal{P}_\mathrm{iPCLR}(\mathbf{x}) ,
\label{eq:pro1}
\end{equation}
where $\mathbf{d}$ is the noisy input signal, and $\lambda > 0$ is a regularization parameter.
Here, the instantaneous phase correction matrix $\mathbf{E}$ for iPC-STFT is fixed to that calculated from the noisy input signal, and thus iPC-STFT is a fixed linear operator.
This simple denoiser is called \textit{proximity operator} which is a building block of the optimization-based signal processing for solving a variety of problems \cite{prox1}.
That is, if Eq.~\eqref{eq:pro1} is effective for denoising, then the proposed prior $\mathcal{P}_\mathrm{iPCLR}$ should be effective in other applications as well.

\vspace{-2pt}
\subsection{Experimental result of audio denoising}
\vspace{-2pt}

The proposed method in Eq.~\eqref{eq:pro1} was applied to audio denoising of the three melodies played by different musical instruments from songKitamura dataset \cite{Kitamura},
where the Gaussian noise was added in the time domain.
The sampling frequency was $44100$ Hz, and STFT was calculated by the canonical tight window of the Hann window of $4096$ samples with $1024$ sample shifting.

The proposed method was compared with other low-rank models: Euclidean NMF (EUC-NMF) for amplitude spectrograms, Itakura--Saito NMF (IS-NMF) \cite{ISNMF}  for power spectrograms, and CNMF\cite{CNMF}.
TSF \cite{TSF} was also compared with a slight modification, the perfect reconstruction constraint $\mathbf{d} = \sum_h \mathbf{x}_h$ was relaxed to a penalty $ \beta/2 \| \mathbf{d} - \sum_h \mathbf{x}_h \|_2^2$, 
because the original formulation of TSF is not suitable for a denoising application.
The number of bases was set to $30$ for the conventional methods (note that the degree of low-rankness is decided by $\lambda$ in the proposed method).
The other parameters of CNMF and TSF were set to the default value in the original papers \cite{CNMF,TSF}.
The number of iterations was set to $100$ for all methods, where ADMM \cite{ADMM} was adopted for solving Eq.~\eqref{eq:pro1}, and $10$ initial values were randomly chosen.
The proposed iPCLR was implemented in two ways for investigating the effect of estimation error on the instantaneous frequency; iPCLRest estimated the instantaneous frequency from the noisy signal, while iPCLRora estimated it from the original clean signal.
In both cases, the estimation was done by Eq.~\eqref{eq:reasign}.
Note that the proposed method and TSF are phase-aware in the sense that the variable is treated in the time domain which ensures consistency \cite{TSF} of the spectrogram.

\begin{figure}[t]
	\centering
	\includegraphics[width=0.96\columnwidth]{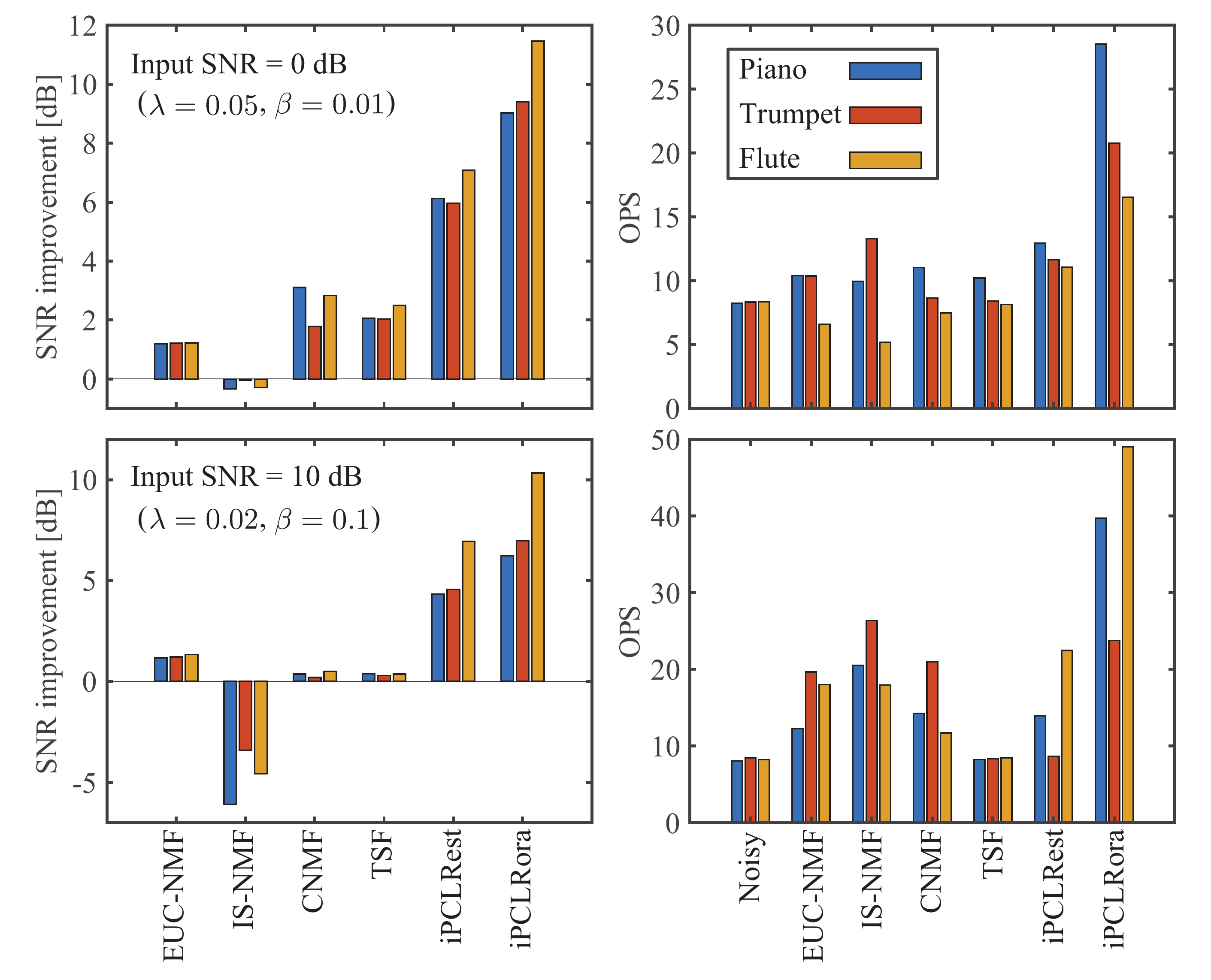}
	\vspace{-10pt}
	\caption{SNR and OPS of the denoising results.
They were the average scores of $10$ trials whose initial values were randomly selected.}
	\label{fig:Resultappl}
	\vspace{-8pt}
\end{figure}

Fig.~\ref{fig:Resultappl} shows the average score calculated from ten random initial values, where the performances were evaluated by SNR and overall-perceptual-score (OPS) which is a perceptual measure available in the PEASS toolbox \cite{PEASS}.
The low-rank models of amplitude spectrograms (EUC- and IS-NMF) obtained limited results as similar to those in the table and figures in the previous section.
This is because those models treat the stationary noise as a part of the low-rank components and do not attempt to reduce it.
For noisier situation (top row), CNMF and TSF obtained higher SNR comparing to EUC- and IS-NMF by considering phase.
On the other hand, their SNR improvement were lower than that of EUC-NMF when the input SNR $= 10$ dB (bottom row), which should be because they do not consider the explicit structure of phase that causes instability.
In contrast to the conventional methods, the proposed method in Eq.~\eqref{eq:pro1} achieved better scores by taking advantage of considering the structure of the phase given by Eq.~\eqref{eq:sinusoidalmodel}, even when the instantaneous frequency was estimated from the noisy observations (iPCLRest).
Thanks to the accurate estimation of the instantaneous frequency, iPCLRora resulted in the highest SNR and OPS.
However, we stress that iPCLRest also worked well in terms of SNR because the instantaneous frequency around the spectral peaks can be accurately estimated.
The error of the instantaneous frequency at the time-frequency bin with small amplitude does not significantly affect to the proposed method.

\vspace{-2pt}
\section{Conclusion}
\vspace{-2pt}

In this paper, we showed that the rank of a complex-valued spectrogram can be as low as its amplitude by applying the instantaneous phase correction under mild assumptions.
Based on this finding, a low-rank model called iPCLR was proposed for audio signal processing, and its potentiality was illustrated through audio denoising.
Seeking further applications of iPCLR remains as future works.

\clearpage
\bibliographystyle{IEEEbib}

\end{document}